\documentclass[12pt]{article}
\usepackage{maple2e}
\DefineParaStyle{Maple Output} \DefineParaStyle{Maple Plot}
\DefineParaStyle{Warning} \DefineCharStyle{2D Math}
\DefineCharStyle{2D Output}
\title{Lorenz integrable system moves \`a la Poinsot}
\author{M. C. Nucci}
\date{Dipartimento di Matematica e Informatica\\
 Universit\`{a} di Perugia,
 06123 Perugia, Italy \\
e-mail: {\tt nucci@unipg.it}}
\begin{document}
\baselineskip=24pt
 \maketitle
\begin{abstract}
A transformation is derived which takes the Lorenz integrable
system into the well-known Euler equations of a torque-free  rigid
body about a fixed point, i.e. the famous motion \`a la Poinsot.
The proof is based on Lie group analysis applied to two
third-order ordinary differential equations admitting the same
two-dimensional Lie symmetry algebra. Lie's classification of
two-dimensional symmetry algebras in the plane is used. If the
same transformation is applied to the Lorenz system with any
values of the parameters, then one obtains Euler equations of a
rigid body about a fixed point subjected to a torsion depending on
time and angular velocity. The numerical solution of this system
yields a three-dimensional picture which looks like a ``tornado"
the cross-section of which has a butterfly-shape. Thus Lorenz's
{\em butterfly} has been transformed into a {\em tornado}.
\end{abstract}
{\bf Running title:} Lorenz integrable system moves \`a la Poinsot
\newpage
\section{Introduction}
The motion of a heavy rigid body about a fixed point is one of the
most famous problems of classical mechanics \cite{Golubev}. In
1750 Euler \cite{Euler} derived the equations of motion, which now
bear his name, and described what is nowadays known as the
Euler-Poinsot case because of  the geometrical description given
by Poinsot about hundred years later \cite{Poinsot}.  It was
Jacobi \cite{Jacobi1} who integrated this case by using the
elliptic functions  which he had developed (along with Legendre,
Abel and Gauss \cite{Natucci}) and  mastered \cite{Jacobi2} -- we
have translated this fundamental text into Italian and commented
extensively \cite{RNM}.
\\
 More than 200 years later, in 1963, a paper was
published \cite{Lorenz} in which was presented a system of three
ordinary differential equations.
 The author considered a hydrodynamical system developed by Rayleigh \cite{Rayleigh}
 and reduced it by applying a double Fourier series as in
 \cite{Saltzman}. Thus he obtained what nowadays is the famous
 Lorenz system \cite{Dalmedico}. Three parameters are part of the Lorenz system.
  For particular  values of those parameters the
 Lorenz system can be integrated in closed form
  by means of Jacobi elliptic
functions \cite{Segur}. We call this system the Lorenz integrable
system.
\\
In January 2001 the first Whiteman prize for notable exposition on
the history of mathematics was awarded to Thomas Hawkins by the
American Mathematical Society. In the citation, published in the
Notices of AMS {\bf 48} 416-417 (2001), one reads that Thomas
Hawkins ``$\ldots$ has written extensively on the history of Lie
groups. In particular he has traced their origins to [Lie's] work
in the 1870s on differential equations $\ldots$ the {\em id\'ee
fixe} guiding Lie's work was the development of a Galois theory of
differential equations $\ldots$ [Hawkins's book \cite{Hawkins2}]
highlights the fascinating interaction of geometry, analysis,
mathematical physics, algebra and topology $\ldots$". Also Hawkins
had established ``the nature and extent of Jacobi's influence upon
Lie" \cite{Hawkins1}.
\\
 In the Introduction of his
book \cite{Olver}  Olver wrote that ``it is impossible to
overestimate the importance of Lie's contribution to modern
science and mathematics. Nevertheless anyone who is already
familiar with [it] $\dots$ is perhaps surprised to know that its
original inspirational source was the field of differential
equations".
\\
 Lie's monumental work on transformation groups,
\cite{Lie 1}, \cite{Lie 2} and \cite{Lie 3}, and in particular
contact transformations \cite{Lie 4}, led him to achieve his goal
\cite{Lie 5}.
\\
 Many books have been dedicated to this subject
and its generalizations (\cite{Ame72}, \cite{BluC74}, \cite{Ovs},
\cite{Olver}, \cite{BluK}, \cite{RogA}, \cite{Ste}, \cite{Hil},
\cite{crc1} and \cite{crc2}).
\\
 Lie group
analysis is indeed the most powerful tool to find the general
solution of ordinary differential equations. Any known integration
technique can be shown to be a particular case of a general
integration method based on the derivation of the continuous group
of symmetries admitted by the differential equation, i.e. the Lie
symmetry algebra. In particular Bianchi's theorem (\cite{Bianchi},
\cite{Olver}) states that, if an admitted $n$-dimensional solvable
Lie symmetry algebra is found, then the general solution of  the
corresponding  $n^{th}$-order system of ordinary differential
equations can be obtained by quadratures. The admitted Lie
symmetry algebra can be easily derived by a straightforward
although lengthy procedure. As  computer algebra software becomes
widely used, the integration of systems of ordinary differential
equations by means of  Lie group analysis is becoming easier to
perform. A major drawback of Lie's method is that it is useless
when applied to systems of $n$ first-order equations, because they
admit an infinite number of symmetries, and there is no systematic
way to find even an one-dimensional Lie symmetry algebra, apart
from trivial groups like translations in time admitted by
autonomous systems. One may try to derive an admitted
$n$-dimensional solvable Lie symmetry algebra by making an ansatz
on the form of its generators. \\However, in \cite{kepler} we have
remarked that any system of $n$ first-order equations could be
transformed into an equivalent system where at least one of the
equations is of second-order. Then the admitted Lie symmetry
algebra is no longer infinite-dimensional, and nontrivial
symmetries of the original system could be retrieved
\cite{kepler}. This idea has been successfully applied in several
instances (\cite{kepler}, \cite{valenu}, \cite{harmony} and
\cite{classic}). Also in \cite{marcelnuc} we have have shown that
first integrals can be obtained by Lie group analysis even if the
system under study does not come from a variational problem, i.e.,
without making use of Noether's theorem \cite{Noether}. If we
consider a system of first-order equations and, by eliminating one
of the dependent variables, derive an equivalent system which has
one equation of second-order, then Lie group analysis applied to
that equivalent system yields the first integral(s)  of the
original system which do(es) not contain the eliminated dependent
variable. Of course this requires that such first integrals exist.
  The procedure should be repeated as many times as there are dependent
  variables  in order to find all such first integrals.
  The first integrals correspond to the characteristic curves of determining equations of
parabolic type which are constructed by the method of Lie group
analysis. We remark that interactive (not automatic) programs for
calculating Lie symmetries such as \cite{man1} and \cite{man2} are
more appropriate for performing this task.
\\
We have  briefly sketched three apparently unrelated subjects. In
this paper we show that the Lorenz system and the Euler equations
are actually related by means of Lie group analysis.
\\
In \cite{issue95} we applied Lie group analysis to a third-order
differential equation, which is equivalent to the Lorenz
integrable system, and obtained a two-dimensional Lie symmetry
algebra, which we then used to integrate the Lorenz integrable
system in terms of Jacobi elliptic functions. Here we show that
the same Lie symmetry algebra is admitted by a third-order
differential equation which is equivalent to the Euler equations
of a torque-free rigid body moving about a fixed point. Then a
transformation is easily derived by which the Lorenz integrable
system becomes the Euler equations of a torque-free rigid body
moving about a fixed point. Thus, it can be stated that ``the
Lorenz integrable system moves \`a la Poinsot".
\\
If the same transformation is applied to the Lorenz system with
any value of parameters, then one obtains the Euler equations of a
rigid body moving about a fixed point and subjected to a torsion
depending on time and angular velocity. The numerical solution of
this system yields a three-dimensional picture which resembles a
``tornado" the cross-section of which has a butterfly-shape. By
means of our transformation Lorenz's {\em butterfly} becomes a
{\em tornado}.
\\
In the last section  the relationship between Lie group analysis
and first integrals \cite{marcelnuc} is exemplified  by
considering the Euler equations of a torque-free rigid body about
a fixed point and the Lorenz integrable system.

\section{Butterflies and tornadoes}

Consider the Lorenz system \cite{Lorenz}:
\begin{eqnarray}
x'&=& \tilde{\sigma} (y-x), \label{lor1} \\ y'&=& -xz+\tilde{r}
x-y, \label{lor2}
\\ z'&=& xy-\tilde{b} z, \label{lor3}
\end{eqnarray}
where $\tilde{\sigma}$, $\tilde{b}$ and $\tilde{r}$ are parameters
(a prime denotes differentiation with respect to $\tau$). This
system can be reduced to a single third-order ordinary
differential equation for $x$ \cite{SenT}, which admits a
two-dimensional Lie symmetry algebra if $\tilde{\sigma}=1/2$,
$\tilde{b}=1$ and $\tilde{r}=0$. System (\ref{lor1}-\ref{lor3})
becomes
\begin{eqnarray}
x'&=&{(y-x)\over 2}, \label{lor1m} \\ y'&=& -xz-y, \label{lor2m}
\\ z'&=& xy- z. \label{lor3m}
\end{eqnarray}
The corresponding third-order equation is:
\begin{equation}
2xx'''-2x'x''+5xx''-3x'^2+2x^3x'+3xx'+x^4+x^2=0, \label{loreq}
\end{equation}
and admits a two-dimensional Lie symmetry algebra $L_2$ with
basis:
\begin{equation}
X_1=\partial_{\tau},\;\;\;\;\;\;\;\;\;
X_2=e^{\tau/2}\left(\partial_{\tau}-{1\over 2}x\partial_x\right).
\label{Xlor}
\end{equation}
 A basis of its differential invariants of-order $\leq 2$ is given by:
\begin{equation}
\phi=\left(x'+{x\over 2}\right)x^{-2},\;\;\;\;\;\;\;\;
\psi=\left(x''+{3\over 2}x'+{x\over 2}\right)x^{-3}.
\label{phipsilor}
\end{equation}
Equation (\ref{loreq}) is reduced to the following first-order
equation:
\begin{equation}
\left(\psi-2\phi^2\right){{\rm d} \psi \over {\rm d}
\phi}=-2\psi\phi-\phi \label{feqlor},
\end{equation}
which can be easily integrated \cite{Murphy} to give:
\begin{equation}
{1+4\psi-4\phi^2\over (1+2\psi)^2}=c_1, \label{psisoll}
\end{equation}
where $c_1$ is an arbitrary constant. Substitution of $x$ and its
derivatives into (\ref{psisoll}) yields a second-order ordinary
differential equation
\begin{equation}
{1+4 \left(x''+{3\over 2}x'+{1\over 2}x\right)x^{-3}
-4\left(x'+{1\over 2}x\right)^2x^{-4} \over
\left(x^3+2x''+3x'+x\right)^2x^{-6} }=c_1, \label{x2ode}
\end{equation}
which admits the Lie symmetry algebra $L_2$. Lie's classification
of two-dimensional algebras into four canonical types \cite{Lie 5}
allows us to integrate (\ref{x2ode}) by quadrature if we introduce
the canonical variables:
\begin{equation}
v=-2e^{-\tau/2},\;\;\;\;\;\;\;\;u={e^{-\tau/2}\over
x},\label{uvlor}
\end{equation}
which transform equation (\ref{x2ode}) into
\begin{equation}
{1+4 \left({\displaystyle{{\rm d} u \over {\rm d}v}}\right)^2-4u
{\displaystyle{{\rm d}^2 u \over {\rm d}v^2}} \over
\left[2u{\displaystyle{{\rm d}^2 u \over {\rm d}v^2}} -4
\left({\displaystyle{{\rm d} u \over {\rm
d}v}}\right)^2-1\right]^2}=c_1, \label{ueqlor}
\end{equation}
and operators (\ref{Xlor}) into
\begin{equation}
\bar X_1=\partial_v,\;\;\;\;\;\;\;\;\; \bar
X_2=v\partial_v+u\partial_u \, . \label{Xblor}
\end{equation}
Then the general solution of (\ref{ueqlor}) can be easily derived
\cite{Lie 5} to be:
\begin{equation}
\displaystyle\int \left(-c_1\mp2c_2u^2-c_2^2u^4\right)^{-1/2}{\rm
d} u=\pm {v \over 2\sqrt{c_1}}+c_3,
\end{equation}
with $c_2$ and $c_3$ arbitrary constants. This solution which
involves an elliptic integral has already been obtained by Sen and
Tabor \cite{SenT} by means of a lengthier analysis. \\
The Euler equations describing the motion of a heavy rigid body
about a fixed point with no torsion are
\begin{eqnarray}
\dot{p}&=& {(B-C)\over A}qr, \label{eul1} \\ \dot q&=& {(C-A)\over
B} pr,\label{eul2}
\\ \dot r&=& {(A-B) \over C}pq, \label{eul3}
\end{eqnarray}
with $A,B$ and $C$ being the principal moments of inertia, and
$p(t),q(t)$ and $r(t)$ the components of the angular velocity (a
dot denotes differentiation with respect to $t$). This system can
be reduced to a single third-order ordinary differential equation
for, say, $p$, viz
\begin{equation}
p{{\rm d}^3 p\over {\rm d} t^3} -{{\rm d} p\over {\rm d} t}{{\rm
d}^2 p\over {\rm d} t^2}-{4(C-A)(A-B)\over B C}p^3 {{\rm d} p\over
{\rm d} t}=0,
\end{equation}
which admits a two-dimensional Lie symmetry algebra ${\cal {L}}_2$
with basis:
\begin{equation}
\Gamma_1=\partial_{t},\;\;\;\;\;\;\;\;\;
\Gamma_2=t\partial_{t}-p\partial_p\,. \label{peul}
\end{equation}
The two Lie symmetry algebras $L_2$ and ${\cal {L}}_2$ that we
have found are actually the same, i.e. Type IV in Lie's
classification \cite{Lie 5}. Therefore they are linked by a
transformation which takes $(\tau,x)$ into $(t,p)$. Prolongation
to the second-order of the two equivalent Lie symmetry algebras
yields a transformation which takes the system
(\ref{eul1})-(\ref{eul3}) into the system
(\ref{lor1m})-(\ref{lor3m}) as
\begin{eqnarray}
\tau&=&\log\left({4\over t^2}\right),\nonumber\\
x&=&{p\,t \over 2},\nonumber\\y&=& {C-B\over 2\,A} q r
t^2,\nonumber\\z&=&  {C-B \over 2A}\left[{(C-A)\over B}
r^2+{(A-B)\over C}q^2\right]t^2,\label{eptolor}
\end{eqnarray}
with the following condition on the momenta of inertia
 \begin{equation}{(A-B)(A-C)\over BC}={1 \over 4}.\label{condo}\end{equation}
 A slightly more general condition could have been considered if one
 replaces $1/4$ with $k/4$ ($k$ an arbitrary parameter).
 If one derives $B$ from (\ref{condo}) by  assuming $A-C>0$ and $4A-3C>0$,
 i.e.
 \begin{equation}
B={4  A  (A - C)\over 4  A - 3  C},\label{bcond}
 \end{equation}
then the transformation (\ref{eptolor}) turns into the following

\begin{eqnarray}
\tau&=&\log\left({4\over t^2}\right),\nonumber\\
x&=&{p\,t \over 2},\nonumber\\y&=& - {(2A-C)(2A-3C)\over 2 A(4A -
3C)}q r t^2, \nonumber\\z&=& -{(2A-C)(2A-3C)\left[4  A^2  q^2-(4
A-3  C)^2 r^2\right]\over 8 A^2(4A - 3C)^2} t^2,\label{eptolorcb}
\end{eqnarray}

and the system (\ref{eul1})-(\ref{eul3}) assumes the form
\begin{eqnarray}
\dot{p}&=& {  (2  A - C)  (2  A - 3  C)  \over A (4 A -3  C)}q r,
\label{eul1sb}
\\ \dot q&=& {3C- 4A\over 4  A} pr, \label{eul2sb}
\\ \dot r&=& {A\over 4A-3C}p q.\label{eul3sb}
\end{eqnarray}
 We also derive the inverse transformation, i.e.
\begin{eqnarray}
t&=&2e^{-\tau/ 2},\nonumber\\
p&=&x e^{\tau/ 2},\nonumber\\
q&=& - {(4A-3C)y e^{\tau/ 2}\over 2\sqrt{(2A-C)(2A-3C)(\sqrt{y^2 + z^2} + z)}},\label{lortoep}\\
r&=&  {A e^{\tau/ 2}\sqrt{\sqrt{y^2 + z^2} + z}\over
\sqrt{(2A-C)(2A-3C)}},\nonumber
\end{eqnarray}
which takes the system (\ref{lor1m})-(\ref{lor3m}) into the system
(\ref{eul1})-(\ref{eul3})  after substituting $B$ as in
(\ref{bcond}).\\
 If one applies the transformation (\ref{lortoep}) to the
general Lorenz  system (\ref{lor1})-(\ref{lor3}), then the
following equations are obtained
%\begin{widetext}
\begin{eqnarray}
\dot{p}&=& {2  (2  A - C)  (2  A - 3  C)  \tilde{\sigma}\over A (4
A -3  C)}q r+(2  \tilde{\sigma} - 1) \, {p\over t},  \label{euln1}
\\ \dot q&=& {3C- 4A\over 4  A} pr +(\tilde{b}-1){4  A^2  q^2-(4  A-3  C)^2  r^2\over
4  A^2  q^2+(4A-3C)^2  r^2}\,{q\over t}\nonumber \\&&+
\tilde{r}{2(4A - 3C)^3A\over (2A-C)(2A-3C)\left[4A^2q^2+(4A-3C)^2
r^2\right]}\,{pr\over t^2},\label{euln2}
\\ \dot r&=& {A\over 4A-3C}p q-(\tilde{b}-1){4  A^2  q^2-(4  A-3  C)^2  r^2\over
4  A^2  q^2+(4A-3C)^2  r^2}\,{r\over t}\nonumber \\&&+
\tilde{r}{8(4A - 3C)A^3 \over (2A-C)(2A-3C)\left[4A^2q^2+(4A-3C)^2
r^2\right]}\,{pq\over t^2}.\label{euln3}
\end{eqnarray}
%\end{widetext}
They can be interpreted as the Euler equations of a rigid body
moving about a fixed point and subjected to a torsion which
depends on time $t$ and angular velocity $(p,q,r)$ in the
body-frame reference. Also the momenta of inertia are linked by
relation (\ref{bcond}).\\
To our knowledge such a system has never been described.\\
 If we use Maple V in order to draw a
three-dimensional plot of system (\ref{euln1})-(\ref{euln3}), then
a ``tornado", the cross-section of which resembles a butterfly, is
obtained (FIG. 1). The usual values for the Lorenz parameters,
$\tilde{\sigma}=10,\,\tilde{b}=8/3,\,\tilde{r}=28$, are imposed.
Also we assume $A=2$ and  $C=1$. We consider $t$ as it varies in
the interval $[2,0.015]$, which corresponds to $\tau\in[0,9.8]$
approximately. The step size used is 0.00005.\\
The butterfly-shape curve is better seen in FIG. 2, which shows
the two-dimensional plot of $p$ versus $q$.\\
 A clearer view of the
``tornado" is given in FIG. 3, which shows the two-dimensional
plot of $r$ versus $q$. Another view can be found in FIG. 4 which
shows the two-dimensional plot of $r$ versus $p$.\\
 In FIG. 5 and 6
the plot of $p$ versus $t$ is given for two different but close
initial values (1 and 1.01). For relatively large values of $t$,
say $t\in[2,0.5]$, the solutions are the same (FIG. 5). For small
$t$, say $t\in  [0.05,0.015]$, a dramatic difference appears (FIG.
6).\\
In \cite{Kus} it was found that  one could explicitly derive a
first integral of the Lorenz system (\ref{lor1})-(\ref{lor3}) in
six different instances which correspond to particular values of
the Lorenz parameters. It is a trivial task to apply those
findings to system (\ref{euln1})-(\ref{euln3}) by using the
transformation (\ref{eptolor}). The following is the list of the
six cases with a relative first integral $F$ for the system
(\ref{euln1})-(\ref{euln3}).
%\begin{widetext}
\begin{description}
\item{(1)} $\tilde b= 2 \tilde\sigma$, $\tilde r$ arbitrary
$$F=4^{2\tilde\sigma-1}t^{2(1-2\tilde\sigma)}\left({p^2}+
{\left[4  A^2  q^2-(4  A-3  C)^2  r^2\right]
(2A-C)(2A-3C)\tilde\sigma \over  A^2(4A - 3C)^2}\right)\, .
$$
\item{(2)} $ \tilde b=0$, $\tilde\sigma={1\over 3}$, $\tilde r$ arbitrary
\begin{eqnarray}
F&=&4^{1/3}t^{-2/3}\left(-p^2\tilde r - 2{(2A-C)(2A-3C)\over
3A(4A-3C)}p q r t-{3\over 16}p^4 t^2\right .\nonumber \\&&\left
.-{(4 A^2 q^2-(4A-3C)^2 r^2)(2A-C)(2A-3C)\over 8 A^2(4A-3C)^2}p^2
t^2+{(2A-C)^2(2A-3C)^2 \over 3 A^2(4A-3C)^2}q^2 r^2 t^2\right)\,
.\nonumber\end{eqnarray}
\item{(3)} $\tilde b= 1$, $\tilde r=0$, $\tilde\sigma$ arbitrary
$$F={(4 A^2 q^2+(4A-3C)^2 r^2)^2(2A-C)^2(2A-3C)^2 \over 4 A^4
(4A-3C)^4}\, .$$
\item{(4)} $ \tilde b=4$, $\tilde\sigma=1$, $\tilde r$ arbitrary
\begin{eqnarray}
F&=&4t^{-6}\left( 16 p^2 \tilde r + 32 \tilde r{(4 A^2
q^2-(4A-3C)^2 r^2)(2A-C)(2A-3C) \over A^2 (4A-3C)^2}-p^4
t^2\right.\nonumber\\&&\left . -2 {(4 A^2 q^2-(4A-3C)^2
r^2)(2A-C)(2A-3C) \over A^2 (4A-3C)^2} p^2t^2 +32 {(2A-C)(2A-3C)
\over A (4A-3C)} p q r t \right .\nonumber \\&&\left .+16
{(2A-C)^2(2A-3C)^2 \over A^2 (4A-3C)^2} q^2 r^2 t^2 -32{(4 A^2
q^2-(4A-3C)^2 r^2)(2A-C)(2A-3C) \over A^2 (4A-3C)^2} \right)\,
.\nonumber
\end{eqnarray}
\item{(5)} $ \tilde b=1$, $\tilde\sigma=1$, $\tilde r$ arbitrary
$$F=-{4 \over t^{2}}p^2\tilde r +{(4 A^2 q^2+(4A-3C)^2
r^2)^2(2A-C)^2(2A-3C)^2 \over 4A^4 (4A-3C)^4}\, .$$
\item{(6)} $\tilde b= 6 \tilde\sigma-2$, $\tilde r=2\tilde\sigma-1$
\begin{eqnarray}
F&=&4^{4 \tilde\sigma}t^{2(1-4\tilde\sigma)}\left(-{ p^4 t^2 \over
64 \tilde\sigma}  -\tilde\sigma{(4 A^2 q^2-(4A-3C)^2
r^2)(2A-C)(2A-3C)\over 32 A^2 (4A-3C)^2} p^2 t^2
 \right. \nonumber
\\&& \left . +{(2\tilde\sigma-1)^2\over 4\tilde\sigma} p^2
+{(2A-C)(2A-3C) (2\tilde\sigma-1)\over 2 A (4A-3C)}p q r t
+{(2A-C)^2(2A-3C)^2 \tilde\sigma\over 4 A^2 (4A-3C)^2}q^2 r^2
t^2\right)\, . \nonumber
\end{eqnarray}
\end{description}
%\end{widetext}

\section{Lie group analysis and first integrals}
In \cite{marcelnuc} we showed the application of Lie group
analysis in order to obtain  first integrals with at least one
missing variable. Consider Euler equations
(\ref{eul1})-(\ref{eul3}). There exist two well-known first
integrals, i.e. the conservation of kinetic energy and the
conservation of angular momentum:
 \begin{equation}
Ap^2+Bq^2+Cr^2=I_1, \label{eulen}
\end{equation}
 \begin{equation}
A^2p^2+B^2q^2+C^2r^2=I_2. \label{eulmom}
\end{equation}
 We derive $p$ from
(\ref{eul3}) (the method works independently of the chosen
equation) as:
$$p={C \dot r\over (A-B)q}$$
in order to obtain the following two differential equations in $q$
and $r$, one of first-order and one second-order, respectively:
\begin{eqnarray}
 \dot q &=&{C (C_A) r  \dot r \over B  (A - B) q} \label{eulu1x}\\
\ddot r&=&  {C(C-A)\over B(A-B)q^2} r\dot r^2+{(B-C) (A-B)\over A
C}r q^2. \label{eulu2xx}\end{eqnarray}

 When  Lie group analysis
of  the system (\ref{eulu1x})-(\ref{eulu2xx}) is performed, a
linear partial differential equation of parabolic structure is
generated. Its characteristic curve is given by
$$B(A-B)q^2+C(A-C)r^2$$ which is a combination of the two first
integrals (\ref{eulen}) and (\ref{eulmom}). Consequently we
introduce the new dependent variable $s$ such that
\begin{equation}
q=\sqrt{{C(C-A)r^2 + s\over B(A-B)}}
\end{equation}
 in order to obtain the following system
 %\begin{widetext}
 \begin{eqnarray}
 \dot s&=&0,\\
 \ddot r&=&{C(C-A)r \dot r^2\over C(C-A)r^2 + s}+{ (C-A)(B-C)r^3\over A
B}+{(B-C)s r\over A B C}.\label{eulfin}
\end{eqnarray}
%\end{widetext}
Equation (\ref{eulfin}) admits an eight-dimensional Lie symmetry
algebra (i.e., it is linearizable) if either $A=C$ or $B=C$, i.e.
the case of the torque-free Lagrange top (uniform precession).\\
 If either $q$ or $r$, one at a time, is eliminated from system (\ref{eul1})-(\ref{eul3}),
  then a similar result is obtained,
i.e. the other two combinations of the first integrals
(\ref{eulen}) and (\ref{eulmom}). Indeed the elimination of $q$
yields the first integral
$$A(A-B)p^2+C(B-C)r^2,$$
and the elimination of  $r$ yields the first integral
$$B(B-C)q^2+A(A-C)p^2.$$
If the same method is applied to the Lorenz integrable system
(\ref{lor1m})-(\ref{lor3m}), then the following two first
integrals are obtained
 \begin{equation}
 (y^2+z^2)\,e^{2\tau}, \label{lormint1}
 \end{equation}
\begin{equation}
(z-x^2)\,e^{\tau}. \label{lormint2}
\end{equation}
 Those first integrals were found in
\cite{Segur} by using the Painlev\'e analysis.

From (\ref{lor3m}) we have
$$x={z'+z\over y}$$
in order to obtain the following two differential equations in $y$
and $z$, one of first-order and one second-order, respectively:
\begin{eqnarray}
  y' &=& - {y^2 + z^2 + z z'\over y},
\label{lormu1x}\\
 z''&=& {y^4 - 3y^2 z - 5y^2 z' - 2z^3 - 4z^2 z' - 2z {z'}^2 \over
 2y^2}.
 \label{lormu2xx}\end{eqnarray}
 When the Lie group analysis
of  the system (\ref{lormu1x})-(\ref{lormu2xx}) is performed, a
linear partial differential equation of parabolic structure is
generated. Its characteristic curve is given by $$y^2+z^2$$
Consequently we introduce the new dependent variable $Y$ such that
\begin{equation}
y=\sqrt{Y-z^2}
\end{equation}
 in order to obtain the following system
 %\begin{widetext}
 \begin{eqnarray}
 Y'&=&-2 Y,\\
 z''&=& {Y^2 - 2 Y z^2 - 3 Y z - 5 Y z' + z^4 + z^3 + z^2 z' - 2
z {z'}^2\over 2(Y - z^2)}. \label{lormfin}
\end{eqnarray}
%\end{widetext}
The first equation can be easily integrated to yield the first
integral (\ref{lormint1}).\\
 If we eliminate $y$ from system (\ref{lor1m})-(\ref{lor3m}),  a
similar procedure provides the first integral (\ref{lormint2}).

\newpage

\begin{figure*}
\mapleplot{figrigid01.eps} \caption{\label{Fig. 1} The 3-dim plot
of the ``tornado".} \end{figure*}

\begin{figure*}
\mapleplot{figrigid05.eps} \caption{\label{Fig. 2} The 2-dim plot
of $p$ vs. $q$.}
\end{figure*}

\begin{figure*}
\mapleplot{figrigid06.eps} \caption{\label{Fig. 3} The 2-dim plot
of $r$ vs. $q$.}
\end{figure*}
\begin{figure*}
\mapleplot{rigids04.eps} \caption{\label{Fig. 4} The 2-dim plot of
$r$ vs. $p$.}
\end{figure*}

\begin{figure*}
\mapleplot{rigids03.eps} \caption{\label{Fig. 5} Varying the
initial condition of $p$ in $t=2$  by $0.01$. The two plots of
$p(t), t\in  [2,0.5]$ are indistinguishable.} \mapleplot{fig6.eps}
\caption{\label{Fig. 6} Varying the initial condition of $p$ in
$t=2$  by $0.01$. There are two different plots of $p(t),t\in
[0.05,0.015]$: the darker line represents the plot of $p(t)$ with
initial condition $p=1$ in $t=2$, and the lighter line the plot of
$p(t)$ with initial condition $p=1.01$ in $t=2$.}
\end{figure*}

\end{document}